\documentclass[aps,prb,onecolumn,superscriptaddress,preprint]{revtex4-1}

\usepackage{graphicx}% Include figure files
\usepackage{dcolumn}% Align table columns on decimal point
\usepackage{bm}% bold math
\usepackage{indentfirst}
\usepackage[greek,english]{babel}

\usepackage{epsfig}
\usepackage{graphics}
\usepackage{booktabs}
\usepackage{amssymb}
\usepackage{amsmath}
\usepackage{amsthm}

\usepackage[latin1]{inputenc}
\usepackage{setspace}
\usepackage{makeidx}

\usepackage[mathscr]{eucal}
\usepackage{mathrsfs}
\usepackage{latexsym}
\usepackage[usenames,dvipsnames,svgnames,table]{xcolor}
\usepackage{textgreek}
\usepackage{float}

\begin{document}

\title{Band offset and gap tuning of tetragonal CuO-SrTiO$_3$ heterojunctions.}

%Tuning the band-offset and gap at the epitaxial heterojunction between tetragonal CuO and SrTiO$_3$

%Band-offset engineering of the CuO-SrTiO$_3$ heterojunction through the stabilization of the tetragonal CuO phase.

%Band-offset and gap evolution at the tetragonal-CuO/SrTiO$_3$ heterojunction.

\author{Giovanni Drera}
\affiliation{I-LAMP and Dipartimento di Matematica e Fisica, Universit\`{a} Cattolica del Sacro Cuore, via dei Musei 41, 25121 Brescia (Italy)}
\author{Alessio Giampietri}
\affiliation{Elettra - Sincrotrone Trieste, km. 163,5-s.s.14 in Area Science Park, 34149 Basovizza, Trieste (Italy)}
\author{Alfredo Febbrari}
\affiliation{I-LAMP and Dipartimento di Matematica e Fisica, Universit\`{a} Cattolica del Sacro Cuore, via dei Musei 41, 25121 Brescia (Italy)}
\author{Maddalena Patrini}
\affiliation{Dipartimento di Fisica, Universit\`{a} degli Studi di Pavia, via Bassi 6, 27100 Pavia (Italy)}
\author{Maria Cristina Mozzati}
\affiliation{Dipartimento di Fisica, Universit\`{a} degli Studi di Pavia, via Bassi 6, 27100 Pavia (Italy)}
\author{Luigi Sangaletti}
\affiliation{I-LAMP and Dipartimento di Matematica e Fisica, Universit\`{a} Cattolica del Sacro Cuore, via dei Musei 41, 25121 Brescia (Italy)}
\date{\today}% It is always \today, today,
             %  but any date may be explicitly specified

\begin{abstract}
In this work we analyze the electronic structure at the junction between a SrTiO$_3$ (001) single crystal and a thin tetragonal CuO layer, grown by off-axis RF-sputtering. A detailed characterization of the film growth, based on atomic force microscopy and X-ray photoelectron diffraction measurements, demonstrates the epitaxial growth. We report several markers of a thickness-dependent modification of the film gap, found on both Cu 2p and valence band spectra; through spectroscopic ellipsometry analysis, we provide a direct proof of a band gap increase of the tetragonal CuO layer (1.57 eV) with respect to the thicker monoclinic CuO layer (1.35 eV). This phenomenon is further discussed in the light of cluster calculations and DFT+U simulations. Finally, we report the full experimental band junction diagram, showing a staggered configuration suitable to charge-separation applications, such as photovoltaics and photocatalisys; this configuration is observed up to very low ($<$3 nm) film thickness due to the gap broadening effect.\end{abstract}

\maketitle

\section{Introduction}
The study of the peculiar properties of oxide heterostructures is a very active research topic in modern experimental physics. This renewed interest is influenced by the recent improvements in epitaxial thin film deposition techniques, which now produce samples with unprecedented quality \cite{KosterBook,Debakanta2015}. For this reason, all-oxide epitaxial heterojunctions are providing an extremely rich playground for the development of devices in the field of spintronics, photovoltaics and photocatalysis.

\begin{figure}[ht!]
\begin{center}
\includegraphics[width=0.48\textwidth]{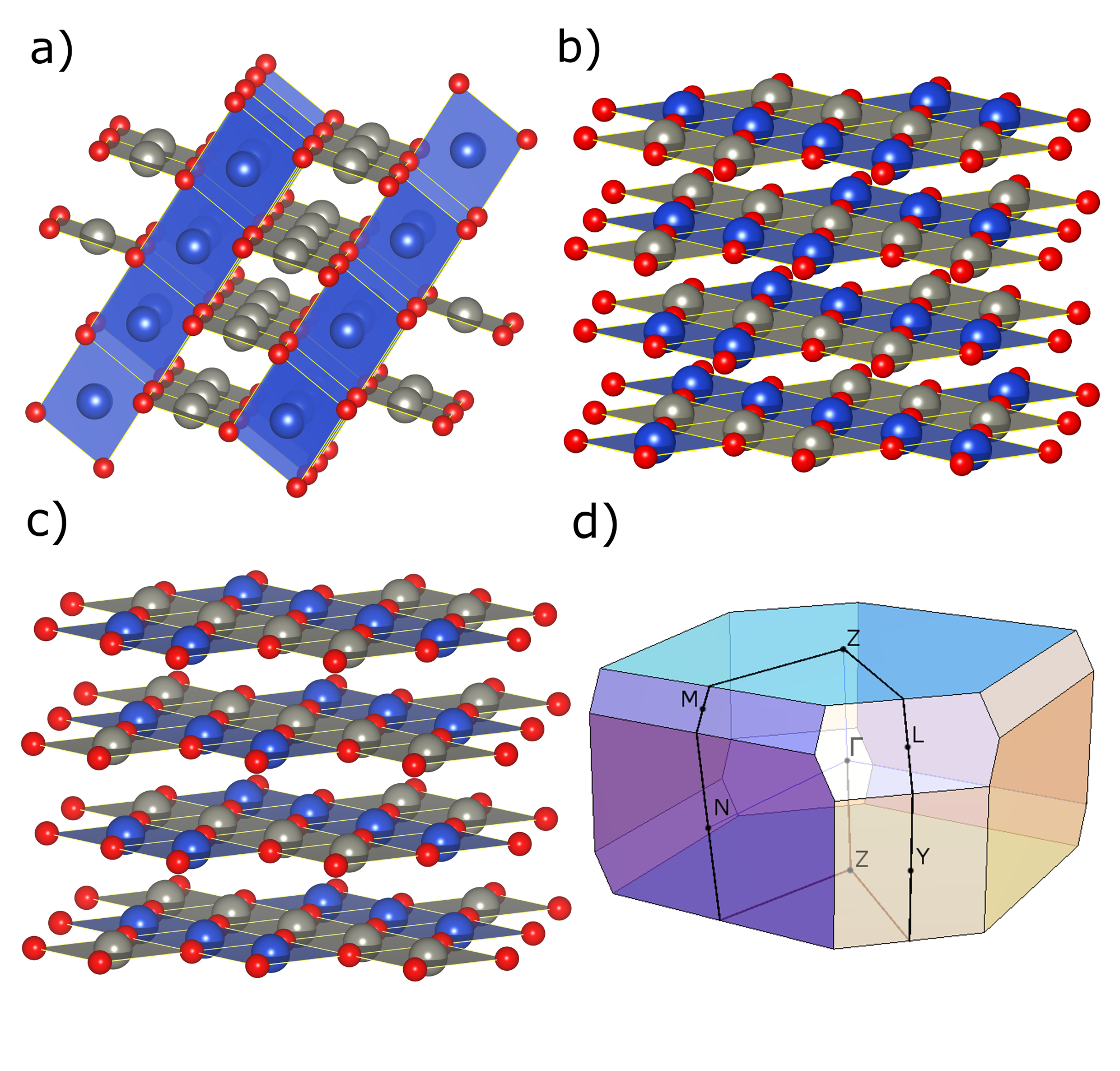}
\caption{Crystal structures for DFT calculation, for m-CuO (a), t-CuO with rocksalt (b) and its nearly degenerate AF order(c). Oxygen atoms are in red, blue and grey are Cu site with opposite spin; the planar CuO$_4$ plaquettes are depicted with the same color of the corresponding Cu atoms. d) Brillouin zone for m-CuO, equivalent to the rocksalt t-CuO one. \label{fig_unitcell}}
\end{center}
\end{figure}

In this context, the cupric oxide (CuO) shows peculiar properties both as a stand alone material and when used in heterojunctions. In fact, CuO is the only late transition metal (TM) monoxide to display a monoclinic unit cell in the bulk phase (hereafter denoted as m-CuO), instead of the usual cubic (rocksalt) structure. While m-CuO shows antiferromagnetism (AF), its N{\'e}el temperature (T$_N$) is considerably lower than expected ($\approx$ 230 K) with respect to other TM oxides. However, a tetragonal CuO phase (t-CuO) has been stabilized in ultrathin epitaxial film grown on SrTiO$_3$ (001) \cite{Siemons8,SamalEurophys2014} (STO). In this phase, the CuO is arranged in a planar structure, with edge-sharing CuO$_4$ square plaquettes (shown in Fig. \ref{fig_unitcell}-b,c), instead of the stripe alignment of m-CuO (Fig. \ref{fig_unitcell}-a).
Calculations\cite{DFT-HSE} predict t-CuO to display AF with T$_N$ up to 900 K; moreover, a Zhang-Rice singlet dispersion similar to high-T$_C$ cuprates has been measured\cite{Grioni2014} by angle-resolved photoelectron spectroscopies, as well as a similar magnon dispersion detected by resonant inelastic x-ray scattering\cite{MoserMagnons2015}. A suitable doping mechanism is still required to exploit a possible superconductivity, which could not be achieved by chemical ways\cite{Grioni2014}.

The CuO/SrTiO$_3$ electronic structure at the interface is also interesting; in fact, such junction belongs to the family of all-oxides STO heterostructures, which are now actively explored since the discovery of the two dimensional electron gas (2DEG) at LaAlO$_3$/SrTiO$_3$ (LAO-STO) interface\cite{Ohtomo15}. This research field has now been expanded to several perovskite and non-perovskite oxide heterojunction\cite{Giampietri17}, such as the $\gamma$-Al$_2$O$_3$/STO\cite{Schutz16} where a 2DEG has been found. In the CuO/STO system, such effect has not yet been investigated. The CuO/STO found also important applications in photoelectrochemical water splitting; in a recent work\cite{Choudhary6}, Choudhary \textit{et al.} demonstrate that the photocatalysis performance of CuO/SrTiO$_3$ heterostructures is increased with respect to that of similar systems, such as ITO/STO and ITO/CuO junctions. CuO/STO also shows a superior photocurrent density and photoconversion efficiency with respect to pristine materials. In order to explain the CuO/STO photocatalysis process, a staggered band junction model has been proposed, in spite of the large difference between the CuO (1.35 eV) and STO gap (3.15 eV).

In this work, CuO thin films are grown on STO by off-axis RF sputtering, with the aim to probe the electronic properties of the junction for different thicknesses (2.7 to 42 nm range) of the CuO overlayer. This allowed us to track the transition from the m-CuO/STO to the t-CuO/STO, which we observed for a thickness below few nm. In particular, the valence band alignment at the junction and the band-gap broadening in t-CuO are investigated, combining indirect (X-ray photoelectron spectroscopy, XPS) and direct (spectroscopic ellipsometry, SE) experimental probes. The results are compared with bulk DFT+U and cluster model calculations, aimed to evaluate the electronic band modification from the monoclinic to the tetragonal phase.

\section{Experimental Details}
The CuO thin films have been grown by RF magnetron sputtering on TiO$_2$-terminated STO (001) substrates, from 2'' polycristalline sputtering targets. In order to properly track the band shift at the interface between two insulators, we avoided the use of Nb-doped STO substrates.
The sputtering power was 80 W, with an Ar flux in the 1.9 - 2.3 sccm range and a 8.5 $\times$ 10 $^{-3}$ mbar pressure. Film crystallization has been achieved with in-growth direct heating at 500$^{\circ}$ C. The STO TiO$_2$ termination has been achieved through HF buffered solution treatment, following the method described by Koster \textit{et al.} \cite{Koster1}. In order to trigger the epitaxial growth, an off-axis deposition geometry has been selected, which reduces the re-sputtering mechanism and allows for a more homogeneous energy distribution of the deposited material \cite{Das3, Baek4}. The growth rate was approximately 0.2 {\AA}/s. Growth morphology analysis has been carried out by atomic force microscopy (AFM), in order to check the STO termination and the CuO film morphology.

The XPS analysis has been performed with the Al K$_\alpha$ line (h$\nu$=1486.6 eV) of a non-monochromatized dual-anode PsP X-ray source and a VG Scienta R3000 electron analyzer operating in transmission mode. This X-ray source has been adopted instead of a monochromatic one in order to mitigate the charging effects due to the insulating STO substrate; in fact, strongly focused X-ray spots, such as the one of monochromatic anode sources or synchrotron radiation, lead to a strong spatial inhomogeneity of the surface charge due to the larger photon flux. Fast repeated acquisitions have been performed for each spectra in order to continuously monitor the relative peak positions, intercalated by several additional spectra on reference C 1s peak; this methodology has been checked by the use of a flood-gun and on a test growth on Nb-doped STO. Adventitious C 1s core level (BE = 284.8 eV) has been used as the energy reference, leading to an overall $\pm$ 0.1 eV absolute error on the binding energy scale.

In X-ray photoelectron diffraction (XPD) measurements the angle between X-ray direction and analyzer axis is fixed at 55.4$^{\circ}$. Single XPD spectra are collected by rotating the sample polar angle ($\theta$) by 5$^{\circ}$ step, exploiting the analyzer angular mode \cite{Drera2}, which allows for the simultaneous acquisition of XPS data in a $\pm$10$^{\circ}$ range in the polar direction. Full stereographic images are collected by performing single XPD spectra acquisition for various azimuthal angles ($\phi$) in the -5$^{\circ}$, 95$^{\circ}$ range with a step of 2.5$^{\circ}$.

Spectroscopic ellipsometry (SE) data have been measured by a VASE spectrometer by J.A. Woollam Inc. in the spectral range from 0.5 to 5 eV at different angles of incidence from 65$^{\circ}$ to 75$^{\circ}$. Experimental spectra were analyzed through the dedicated WVASE 32\textsuperscript{\textcopyright} software and database.

\section{DFT+U calculations}

\begin{figure}[ht!]
\begin{center}
\includegraphics[width=0.46\textwidth]{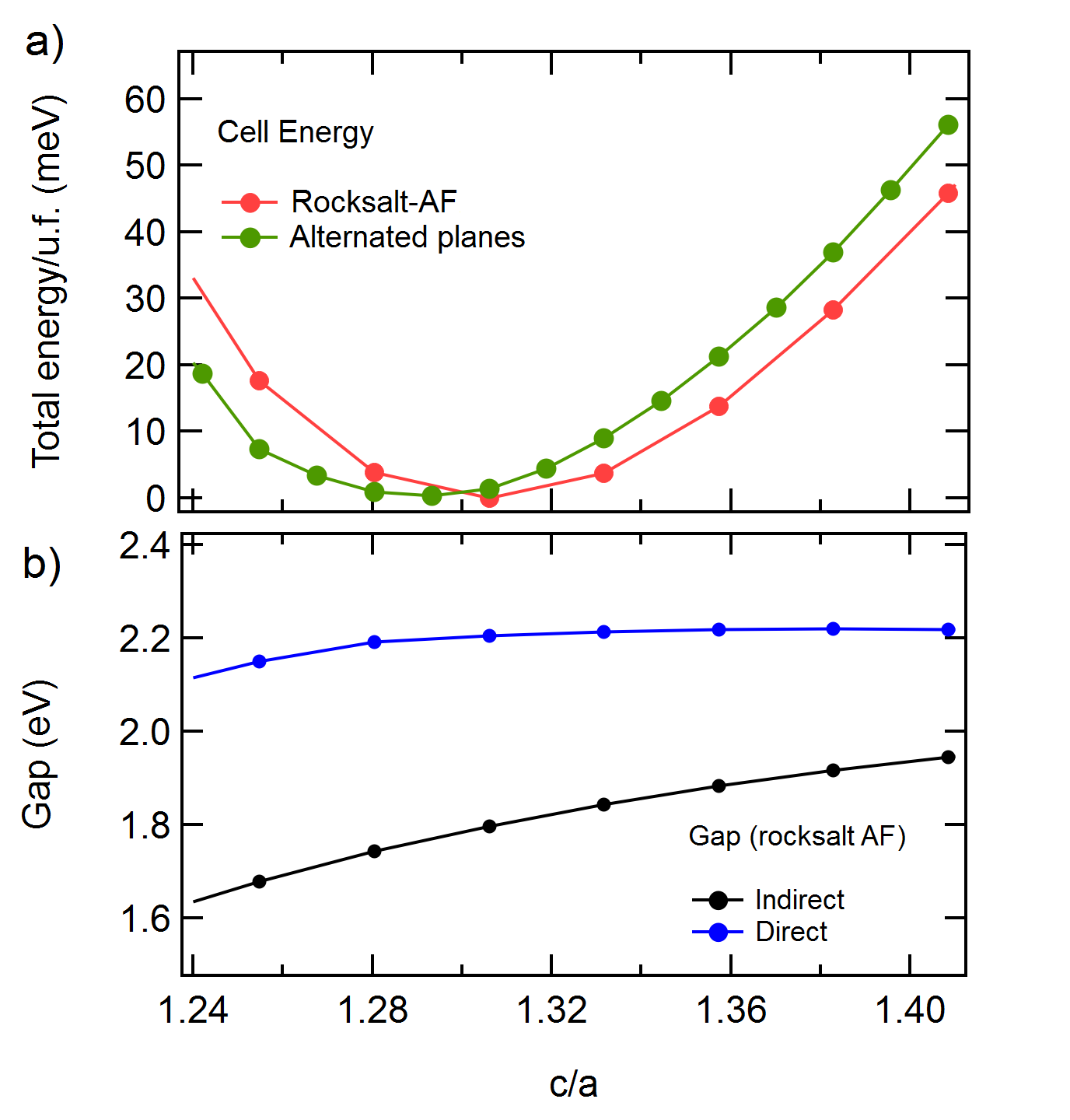}
\caption{a) Calculated total energy per unit formula for the two competing AF order of t-CuO; b) Kohn-Sham direct (blue) and indirect (black) gap for the rocksalt-like AF order. \label{fig_cell_relax}}
\end{center}
\end{figure}

The electronic structure of CuO is notoriously difficult to calculate theoretically with simple ab-initio methods, due to the strong electronic correlation effects. As shown in the next Section, the photoelectron spectroscopy results can be fairly predicted through a cluster model, which takes into account charge-transfer effects and core-hole interaction; however, such approach, as in the case of model Hamiltonians (t-J or Hubbard models) requires the introduction of several free parameters which must be tuned to fit the experimental data. Conversely, a DFT, ab-initio approach allows for the evaluation of a realistic ground-state electronic density, resulting in crystal structures which are often in good agreement with experimental results, and thus to evaluate the effect of actual Cu and O orbital geometry on the band structure.

In spite of the large number of theoretical calculations on CuO, there is still a partial disagreement about the band gap size and type (see, for instance, the review of Meyer et al. \cite{MeyerCuOReviee}). These discrepancies can be attributed to several reasons: the adoption of a non-primitive unit cell, thus resulting in various Brillouin zone topologies; the choice of various AF order vectors; the different theoretical approximations (LDA, LDA+U, hybrid potentials etc...) and computational schemes. For instance, while most of experimental and theoretical works reported monoclinic CuO as an indirect band gap material\cite{WangGWCalc_2016}, some authors reported a direct gap\cite{Ellips_JPN}, or just show a direct gap band structure plot without further specifications\cite{AkumaAnisimovEurPhysJ}.

For these reasons, we performed ab-initio electronic structure calculations on the CuO bulk unit cells, through density functional theory in the framework of spin-resolved LDA-PW approximation\cite{DFT-libxc}. Additional electronic correlation has been introduced in the LDA+U formalism\cite{DFT-DFTU}, since simple LDA results in a metallic ground state; several U values have been considered, namely U$=$5 eV, U$=$6 eV and the ab-initio predicted\cite{DFT-Uvalue} U$=$7.15 eV.

Calculations have been performed with the ABINIT package\cite{DFT-Abinit}, in the framework of Projector-Augmented wave atomic description. A 18 $\times$ 18 $\times$ 18 Monkhorts-Pack grid of points in the reciprocal space and an energy cutoff of 30 Hartree for the plane-wave basis definition have been used for the calculations. The m-CuO cell size has been relaxed up to a maximum interatomic force of $5\cdot 10^{-6}$ Ha/Bohr, leading to a small contraction (1\%) of cell parameter with respect to experimental data for any considered U value. The cell geometry and the spin order are shown in Fig.\ref{fig_unitcell}-a, where each CuO$_4$ plaquette is colored according to the spin sign. The t-CuO cell has been relaxed by keeping a fixed tetragonal base size of 3.905 {\AA}.

\begin{figure*}[ht!]
\begin{center}
\includegraphics[width=0.90\textwidth]{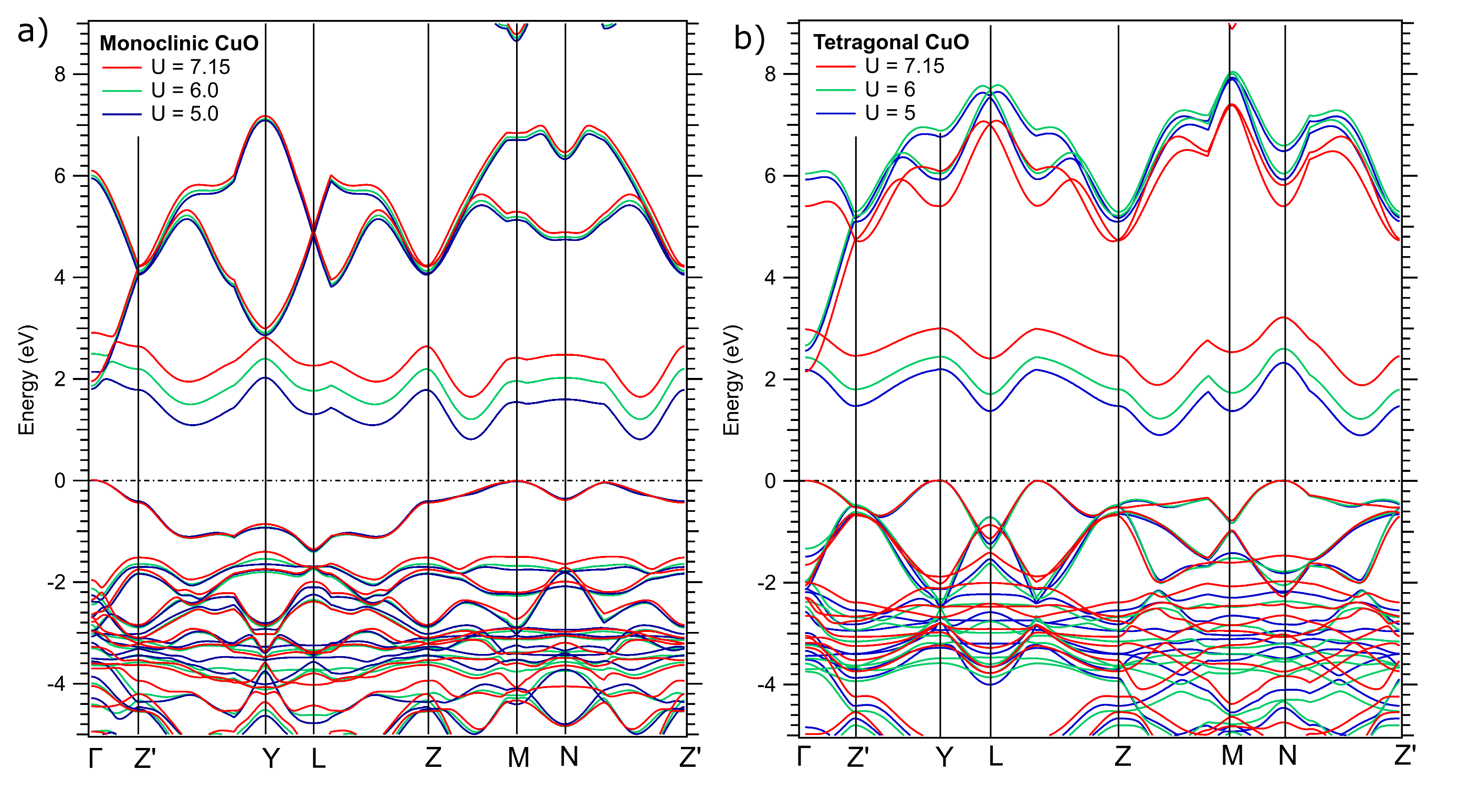}
\caption{Calculated DFT+U band structures for m-CuO (a) and t-CuO(b) for different U values. The band structure path is given in Fig.\ref{fig_unitcell}-d. \label{fig_DFTBands}}
\end{center}
\end{figure*}

For the tetragonal case, several magnetic orders have been considered. The lowest energy was found for two nearly degenerate configurations, with stripes of edge-sharing plaquettes with alternate spin in each plane, shown in Fig.\ref{fig_unitcell}-b,c. For U = 7.15 eV the total energy dependence on $c/a$ ratio is shown by the red and green traces in Fig. \ref{fig_cell_relax}-a. Amongst the t-CuO AF configurations, the absolute lowest energy for $c/a>1$ is a type III AF order with q = (1,1,1) (in cubic conventional coordinates, shown in Fig.\ref{fig_unitcell}-b), which is similar to the usual rocksalt late TM-O structure. In this case, the energy minimum has been found for a tetragonal \emph{c} axis equal to 5.2 {\AA}, corresponding to $c/a=1.33$ which is in excellent agreement with the experimental data\cite{Siemons8}. The second lowest energy structure, shown in Fig. \ref{fig_unitcell}-c, shows a total energy per unit formula just 1 meV higher, although in this configuration the relaxed cell parameter is $c/a = 1.29$, thus lower than the experimentally observed one. The other magnetic order configurations we considered (such as horizontal or vertical CuO planes with alternated spin, not shown in this work) always result in larger total energies (with $\Delta E >$ 200 meV) with different $c/a$ ratios. Our results are similar to hybrid-functional HSE-DFT\cite{DFT-HSE,Franchini7} results. Both lowest energy AF structures show an indirect band gap and very similar spin-resolved electronic densities around Cu atoms.

We then focused on the t-CuO calculated ground state (rocksalt-like); in this case the primitive t-CuO cell becomes monoclinic (C2/m, space group no. 12) as in the m-CuO case (C2/c, space group no. 15), with an AF magnetic order in the (1,0,-2) direction. This particular spin structure is expected only for the hypothetical bulk t-CuO; in fact, in films with a finite thickness the magnetic order can be more complex, as reported by Moser et al.\cite{MoserMagnons2015}. Both t-CuO and m-CuO primitive unit cells consist of two unit formulas, with two inequivalent Cu atoms.

\begin{table}[ht!]
\normalsize
\centering
\begin{tabular}{ccccccc}
              &          &          &m-CuO      &          &              &     \\
              & d$_{xy}$ & d$_{yz}$ & d$_{z^2}$ & d$_{zx}$ & d$_{x^2-y^2}$& Sum \\[2pt]
\hline
 $\uparrow$   & 0.947  & 0.935  & 0.904  & 0.937 & 0.981 & 4.704 \\
 $\downarrow$ & 0.943  & 0.941  & 0.915  & 0.943 & 0.265 & 4.006 \\
 $\uparrow-\downarrow$   & 0.004  & -0.006 & -0.01 &-0.006 & 0.716 & 0.699\\
 \hline
      \\
              &          &          &t-CuO      &          &              &      \\
               & d$_{xy}$ & d$_{yz}$ & d$_{z^2}$ & d$_{zx}$ & d$_{x^2-y^2}$& Sum \\[2pt]
 \hline
 $\uparrow$   & 0.948  & 0.940  & 0.932  & 0.940 & 0.973 & 4.734 \\
 $\downarrow$ & 0.952  & 0.943  & 0.936  & 0.943 & 0.221 & 4.006 \\
 $\uparrow-\downarrow$   & -0.004 & -0.003 & -0.004 &-0.003 & 0.752 & 0.738\\
 \hline
\bottomrule
\end{tabular}
\caption{Orbital occupancy for spin-up and spin-down 3d orbitals of Cu atoms, as calculated by LSDA+U with U = 7.15 eV. The total difference corresponds to the magnetic moment in Bohr magneton units.} \label{tab_occupancy}
\end{table}

The 3d orbital occupancy, as described by the occupation matrix of LSDA+U \cite{DFTU_Zaanen} in real spherical harmonics, is given in Tab. \ref{tab_occupancy} for the U = 7.15 eV case. The average occupation of 3d level for t-CuO and m-CuO is 8.7 e$^-$, consistently with the expected 3d$^9$ configuration of the ionic Cu$^{2+}$ O$^{2-}$ picture. For these calculations, the CuO$_2$ plaquettes have been oriented in order to have Cu-O bond on the x axis direction; as a result, the occupancy differences are nearly completely located on the 3d$_{x^2-y^2}$ orbital. This is true even for m-CuO, where the calculated O-Cu-O angle in the plaquette plane is not exactly 90$^\circ$ ($\approx 83^\circ$). The spin-up and spin-down occupancy difference directly gives the magnetic moment for each Cu atoms (0.70 and 0.74 Bohr magnetons for m-CuO and t-CuO, respectively, for  U = 7.15 eV), which is consistent with other LSDA+U calculations in literature\cite{WangGWCalc_2016}.

Indirect and direct gaps increase with the $c/a$ ratio (Fig. \ref{fig_cell_relax}-b), i.e. with a larger interplanar spacing. Although from the ARPES point of view it is possible to describe the system as a set of weakly interacting planes\cite{Grioni2014}, the orbital hybridization of O 2p with Cu d levels is thus the driving force for the gap modification. With a well relaxed geometry, the Kohn-Sham gap (listed in Tab. \ref{tab_Gap}) of t-CuO is systematically larger respect to m-CuO, for each considered U value. The calculated difference is 0.15 eV for the ab-initio predicted U $=$ 7.15. An indirect fundamental gap is always found, consistently with results in literature \cite{WangGWCalc_2016} and optical spectroscopy measurements\cite{Marabelli19} for m-CuO. At relaxed position, the unit cell volume of t-CuO (38.885 {\AA}$^3$) is quite similar to the m-CuO (39.251{\AA}$^3$), as well as the in-plane plaquette bond length; we thus ascribe the gap change to the relative oxygen orbitals ordering, due to the different plaquettes alignment. While more refined computational schemes could be applied, such as hybrid pseudopotential or dynamical mean field theory, DFT+U can then describe the overall orbital reorganization of CuO in different crystal phases. In any case, these bulk crystal calculations are suitable only for adequately thick films (at least 5/6 u.c., as in this work), since an additional gap change may occur a the interface due to the sharing of oxygen atoms between CuO and SrTiO$_3$, as shown by DFT calculation by Franchini et al.\cite{Franchini7}. However, in order to make these computations feasible, a ferromagnetic ordering was assumed, due to the high computational effort required to carry out calculations with AFM ordering.

The detailed band structures are shown in Fig.\ref{fig_DFTBands}; for the t-CuO case (Fig.\ref{fig_DFTBands}-b), the band structures have been calculated with the relaxed $c/a$ ratios, which are slightly different for each considered U. The Brillouin zone and the interpolation path for band dispersion calculation are shown in Fig.\ref{fig_unitcell}-d; we choose this specific path, which cover all faces middle-points, because it intersects the conduction band minimum (CBM) on the $Z$ point centered face. The Brillouin zones of t-CuO and m-CuO are topologically equivalent. The conduction band minimum is found in a reciprocal space position which is different with respect to previous results for t-CuO (usually identified by the M point\cite{MeyerCuOReviee}). However, as already pointed out, we find several discordant results in literature, due to the adoption of a non-primitive unit cell\cite{DFT-AllCuOxides}, different magnetic structures and different notations for the band structure dispersion.

\begin{table}[ht!]
\normalsize
\centering
\begin{tabular}{lccccccc}
&&m-CuO&&&&t-CuO&\\[2pt]
U (eV)& 5.0 & 6.0 & 7.15 &$\quad$ & 5.0 & 6.0 & 7.15\\
\hline
Indirect Gap  $\quad$      & 0.80 & 1.19 & 1.63 &$\quad$ & 0.88 & 1.23 & 1.78\\
Direct gap $\quad$& 1.06 & 1.43 & 1.86 &$\quad$ & 1.36 & 1.52 & 2.20\\
\bottomrule
\end{tabular}
\caption{DFT+U Kohn-Sham fundamental (indirect) and direct band gap results for m-CuO and t-CuO; all values are in eV.} \label{tab_Gap}
\end{table}

\section{Experiments and discussion}
\subsection{Growth characterization}
\begin{figure}[h!]
\begin{center}
\includegraphics[width=0.49\textwidth]{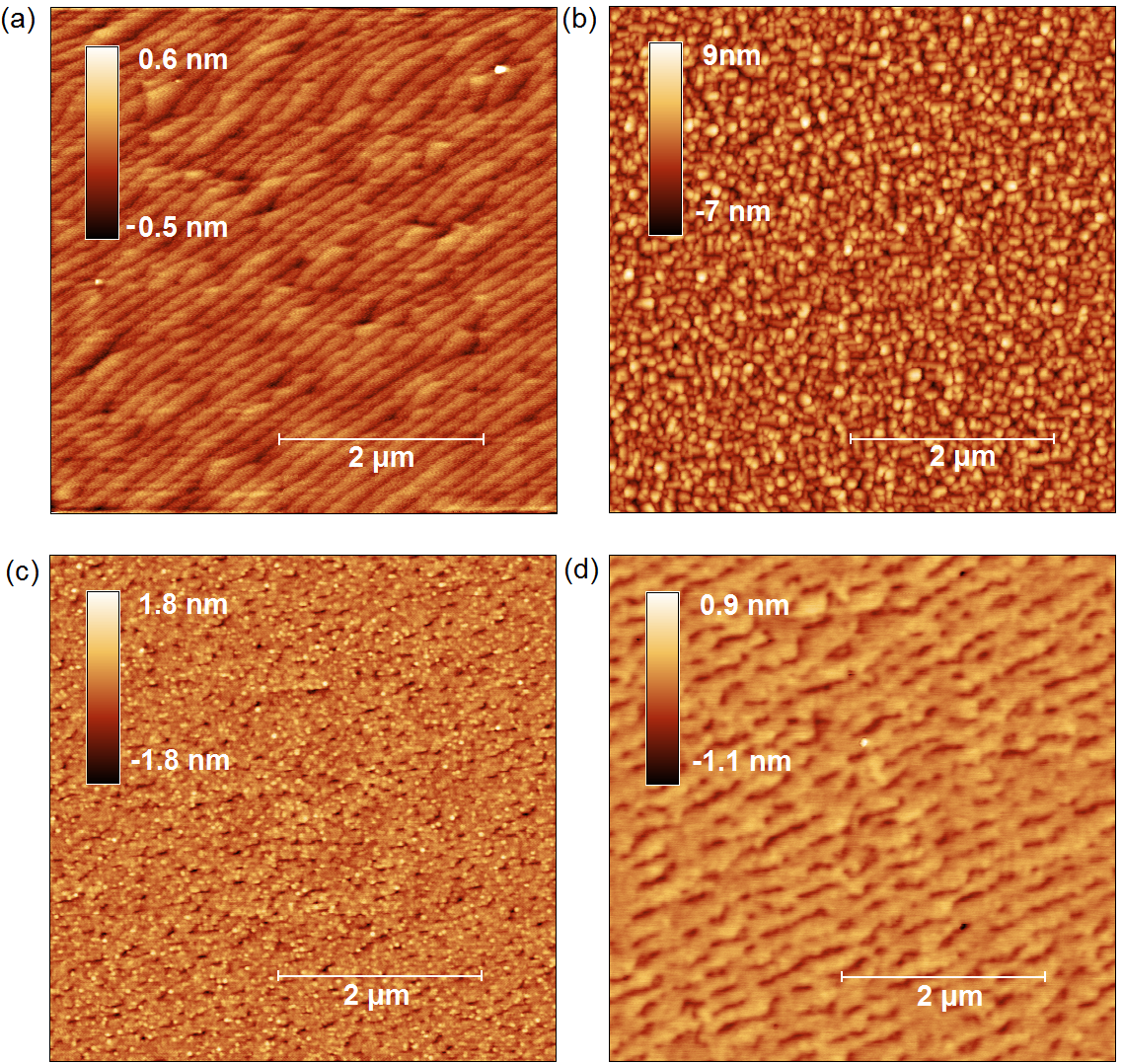}
\caption{AFM topographic analysis carried out on (a) SrTiO$_3$ substrate and on CuO films with decreasing film thickness: (b) 42.0 nm, (c) 4.3 nm, (d) 2.7 nm. Each image covers an area of 5$\times$5 $\mu m$. \label{fig_AFMCuO}}
\end{center}
\end{figure}

In order to obtain high-quality CuO/SrTiO$_3$ heterostructure we initially calibrated the deposition temperature through the growth of CuO films on silicon substrates, and then we grew a set of CuO/SrTiO$_3$ heterostructures with different CuO film thickness (estimated by XPS analysis calibration). The AFM measurements on the thicker film (42 nm, labeled `Thick', Fig.\ref{fig_AFMCuO}-b) shows the presence of a polycrystalline surface. The measured average roughness and crystal size are reduced for thinner films; in fact, the 4.3 nm thick film (hereafter labeled `Thin', Fig.\ref{fig_AFMCuO}-c) displays smaller crystals which start to displace homogeneously along the STO terraces, while an epitaxial CuO layer is obtained for the 2.7 nm thick film (hereafter labeled `Ultrathin', Fig.\ref{fig_AFMCuO}-d). Indeed, due to the strong differences with respect to bulk phase, a stable t-CuO is expected only for a relatively low film thickness.

\begin{table*}[ht!]
\small
\centering
\begin{tabular}{cccccccccc}
Label&Thickness&Thickness&Roughness&Crystal Size&$I_A/I_B$&$\Delta E$&$\Delta$&Q&band-gap\\
&XPS&SE&($\pm$ 0.01 nm)&($\pm$ 0.1 nm)&&(eV)&(eV)&(eV)&SE\\[2pt]
\hline
Thick&42.0&40&0.85&59&2.3&8.07&1.28&8.00&1.35\\
Thin&4.3&4.3&0.18&11&1.9&7.81&1.43&7.98&1.44\\
Ultrathin&2.7&2.79&0.09&-&1.7&7.71&1.47&8.02&1.57\\
\hline
\end{tabular}
\caption{Sample list of t-CuO films grown on SrTiO$_3$ considered in the data analysis, with a summary of the main characterization results. Roughness and crystal size have been measured by AFM; $I_A$, $I_B$, $\Delta E$, $\Delta$ and Q are obtained from the analysis Cu 2p XPS (see text below); the band-gap is measured by SE experiment.} \label{tab_CuO}
\end{table*}

A summary of the film parameters is reported in Tab. \ref{tab_CuO}. The average roughness and crystallites average lateral size deduced from AFM are summarized. For the Ultrathin (2.7 nm) sample grown on SrTiO$_3$ the crystal size is not reported since the film smoothly covers the SrTiO$_3$ terraces.

\begin{figure}[h!]
\begin{center}
\includegraphics[width=0.48\textwidth]{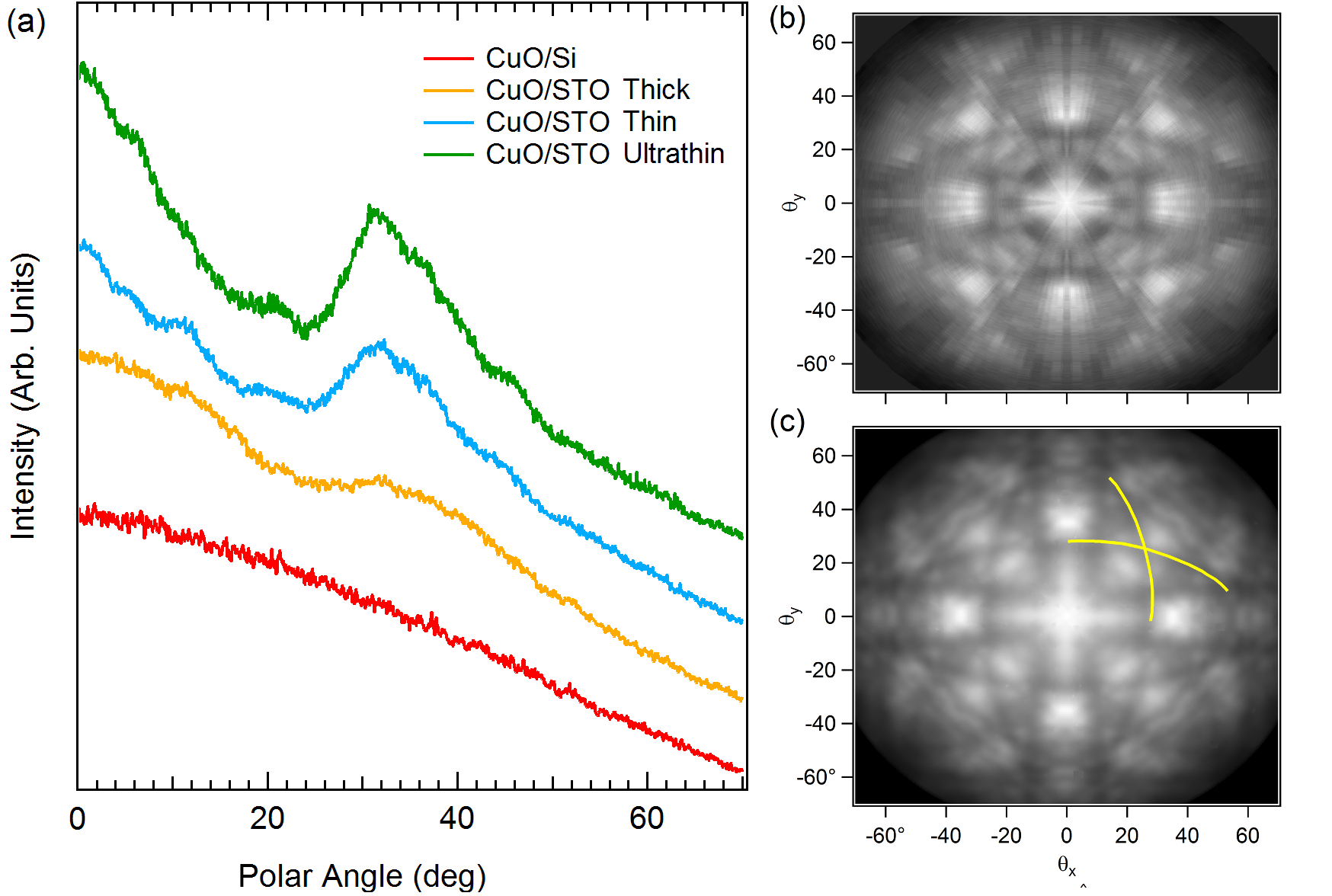}
\caption{(a) XPD analysis of Cu 2p XPS peak area from a CuO/SrTiO$_3$
heterostructure for different film thicknesses. The spectra are collected by sweeping the sample polar angle in the (010) substrate plane, with a fixed azimuthal angle ($\varphi = 0^{\circ}$). (b) Full stereographic images obtained from the Cu 2p XPS peak analysis of Ultrathin film (background subtracted). (c) Multiple scattering simulation for tetragonal Cu 2p, evaluated for $c/a$=1.32 . \label{fig_XPDCuO}}
\end{center}
\end{figure}

The XPS analysis confirms the stoichiometry of the grown films, and the study of the Ti 2p peaks (not shown here) suggests a completely oxidized Ti$^{4+}$ interface, with no residual Ti$^{3+}$ state. The XPS data have also been used to calculate the thickness (and thus the effective deposition rate) through the evaluation of Cu 3p and Sr 3d core level areas.

\begin{figure}[h!]
\begin{center}
\includegraphics[width=0.49\textwidth]{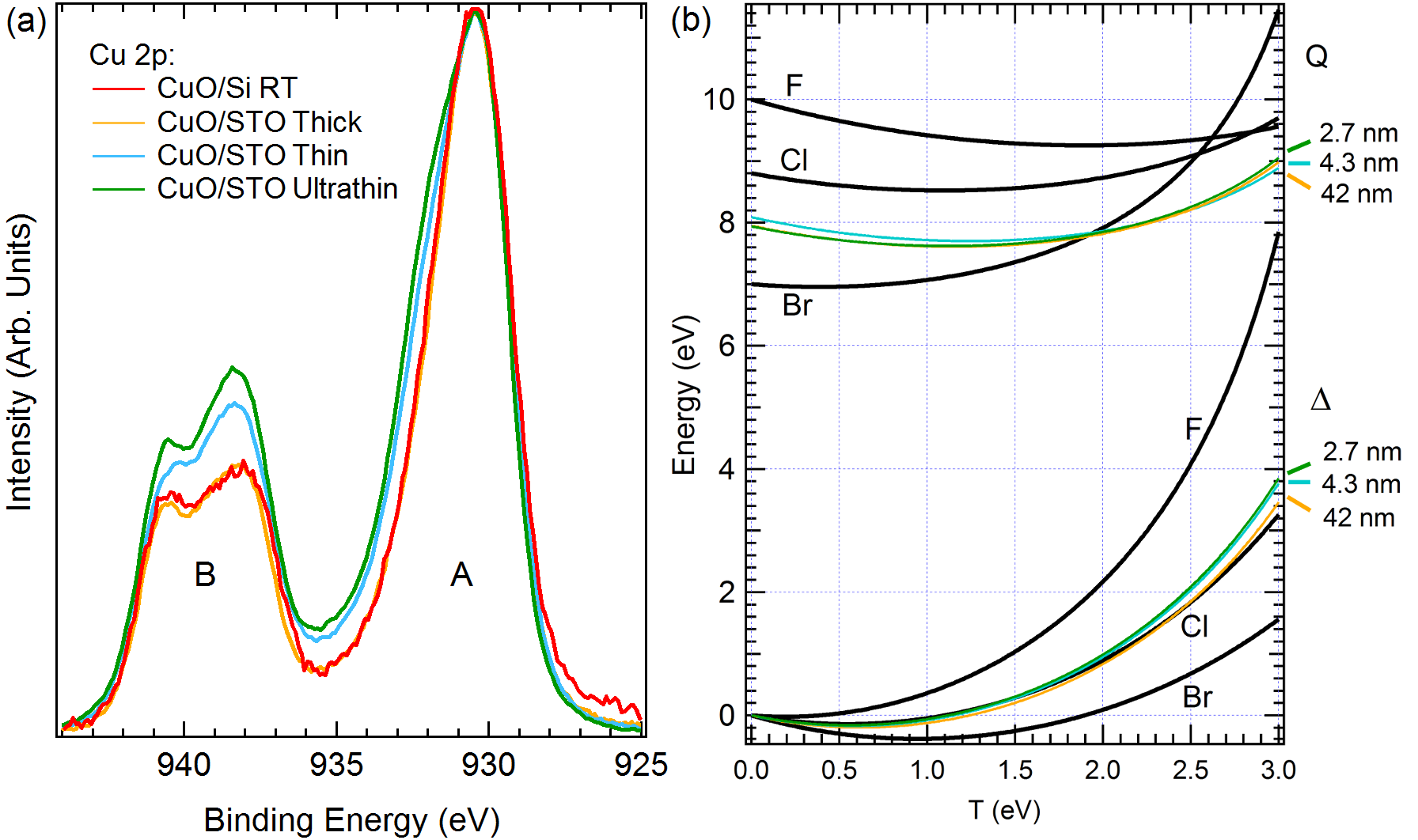}
\caption{(a) Cu 2p$_{3/2}$ XPS peak spectra, collected from films grown on SrTiO$_3$ substrates with different thickness. For quantitative details see Table \ref{tab_CuO}. (b) Variation of the Coulomb interaction $Q$ and of the energy distance between unhybridized one-hole states ($\Delta$) respect to the mixing matrix element of the Hamiltonian ($T$), plotted using experimental values of $\Delta E$ and $I_A/I_B$. Values obtained from copper dihalides, plotted in black, are adapted from Ref. \cite{Vanderlaan13}. Values for Ultrathin, Thin and Thick samples are plotted in green, cyan and orange, respectively. \label{fig_XPSTQDelta}}
\end{center}
\end{figure}

The epitaxial order of CuO films is probed by X-ray photoelectron diffraction (XPD) measurements (Figure \ref{fig_XPDCuO}-(a)). The XPD spectrum of a thick film grown on Si at 600$^{\circ}$C shows a featureless background, produced by the intensity attenuation of XPS peaks area on varying the tilt angle. The growth on a SrTiO$_3$ substrate triggers the appearance of XPD peaks, even for the Thick sample, which are less intense due to the higher roughness (see Table \ref{tab_CuO}). This result, combined with the AFM images, demonstrates the deposition of high-quality and epitaxial CuO films by RF sputtering.

Full stereographic images have been acquired both for Cu 2p and Sr 3d XPS peaks on Ultrathin film, to gain information on the film and the substrate, respectively. In order to prove the crystal order, we also performed multiple scattering simulations through EDAC code\cite{EDAC} on a nearly 700 atom cluster, in order to evaluate the amount of tetragonal distortion from the Cu 2p image (data and simulations are shown in Fig. \ref{fig_XPDCuO} b-c, respectively). The calculation shows a nice agreement with the data; in particular, the dark Kikuchi lines (highlighted in yellow in Fig.\ref{fig_XPDCuO}-c), which crosses at the position of forward scattering peak in the tetragonal (1,1,1) direction, are also reproduced by calculation and can be used to directly estimate a $c/a=1.32$ ratio, in agreement with the previously reported results and with our calculations.

\subsection{Electronic structure}

The deposition of CuO on STO at 500$^{\circ}$C results in a Cu 2p XPS spectrum shape and peak intensity similar to the one observed for a film grown on Si (Fig. \ref{fig_XPSTQDelta}-a). Along with a main line (A) at BE $\approx$ 930.5 eV, the Cu 2p $_{3/2}$ XPS spectra show a characteristic strong satellite (B) at about 939 eV, which is absent in the Cu$^{1+}$, Cu$_2$O compound\cite{IOP2012}. This is a clear indication that the copper oxide films have been grown with the proper oxidation state (Cu$^{2+}$) of the copper ion. However, the satellite intensity and its relative distance from the main line (Table \ref{tab_CuO}) show a clear correlation with the film thickness. Namely, the I$_A$/I$_B$ intensity ratio decreases as the CuO film thickness is reduced.

The measured I$_A$/I$_B$ intensity ratio and the A-B peak binding energy separation ($\Delta E$) of the Cu 2p$_{3/2}$ XPS spectrum can be estimated through configuration-interaction cluster model calculations, i.e. by considering an atomic Cu photoemission site within a coordination cage of oxygen atoms. In this approach, the Cu electronic structure is described as a superposition of 3d$^9$ and 3d$^{10}$\underline{L} configurations, where \underline{L} represents a hole on the oxygen 2p orbital; the energy of each level can then be described through a limited set of parameters, such as the O 2p-Cu 3d charge-transfer $\Delta_{CT}$, the Q$_{pd}$ Coulomb energy, related to the interaction between the Cu 2p core hole created upon photoemission and the d electron in the outer shell of the Cu cation, and the T$_{pd}$ hybridization energy between O 2p and Cu 3d orbitals. The full description of this model can be found in literature\cite{Vanderlaan13}.

Using as input the experimental values of $\Delta E$ and I$_A$/I$_B$, the Q$_{pd}$ vs T and $\Delta_{CT}$ vs T curves can be calculated, as shown in Fig. \ref{fig_XPSTQDelta}-b, where the curves for the present samples have been added to those previously estimated by Van Der Laan et al.\cite{Vanderlaan13} for Cu dihalides and CuO. It is important to note that the T$_{pd}$ value considered in this model should be assumed as an effective value, as in the simple model calculations the effects of the D$_{4h}$ symmetry on T$_{pd}$ are not accounted for.

Unlike the case of Cu halides our Q$_{pd}$ curves are rather flat but, following the procedure of J. Ghijsen et al.\cite{CuOCuO2Ghijsen}, we set a common value for T$_{pd}$ at 2.25 eV, rather similar to the T$_{pd}$ value for bulk CuO at 2.5 eV suggested by Ghijsen \emph{et al.} . According to the Q$_{pd}$ value we assumed, the resulting $\Delta_{CT}$ displays an overall increase of 0.2 eV (i.e. from 1.28 to 1.47 eV) as the film thickness decreases. This ordering does not change in a relatively wide range of T$_{pd}$ energies compatible with the case of cuprates (e.g. in the 2.0-3.0 eV energy range).

Consistently, the sequence of $\Delta_{CT}$ values we report in Table \ref{tab_CuO} is assumed to be an indication of the overall trend of $\Delta_{CT}$ changes with thickness. As these cuprates are classified as CT insulators \cite{CTGAPS_PhysRevLett.55.418,RevModPhys.70.1039}, the increase of charge transfer $\Delta_{CT}$ is expected to ultimately determine an increase of the energy gap, as is consistently found in the energy gap trend measured by SE (Section IV-C) and retrieved from ab-initio band structure calculations (Section III).

\begin{figure}[h!]
\begin{center}
\includegraphics[width=0.50\textwidth]{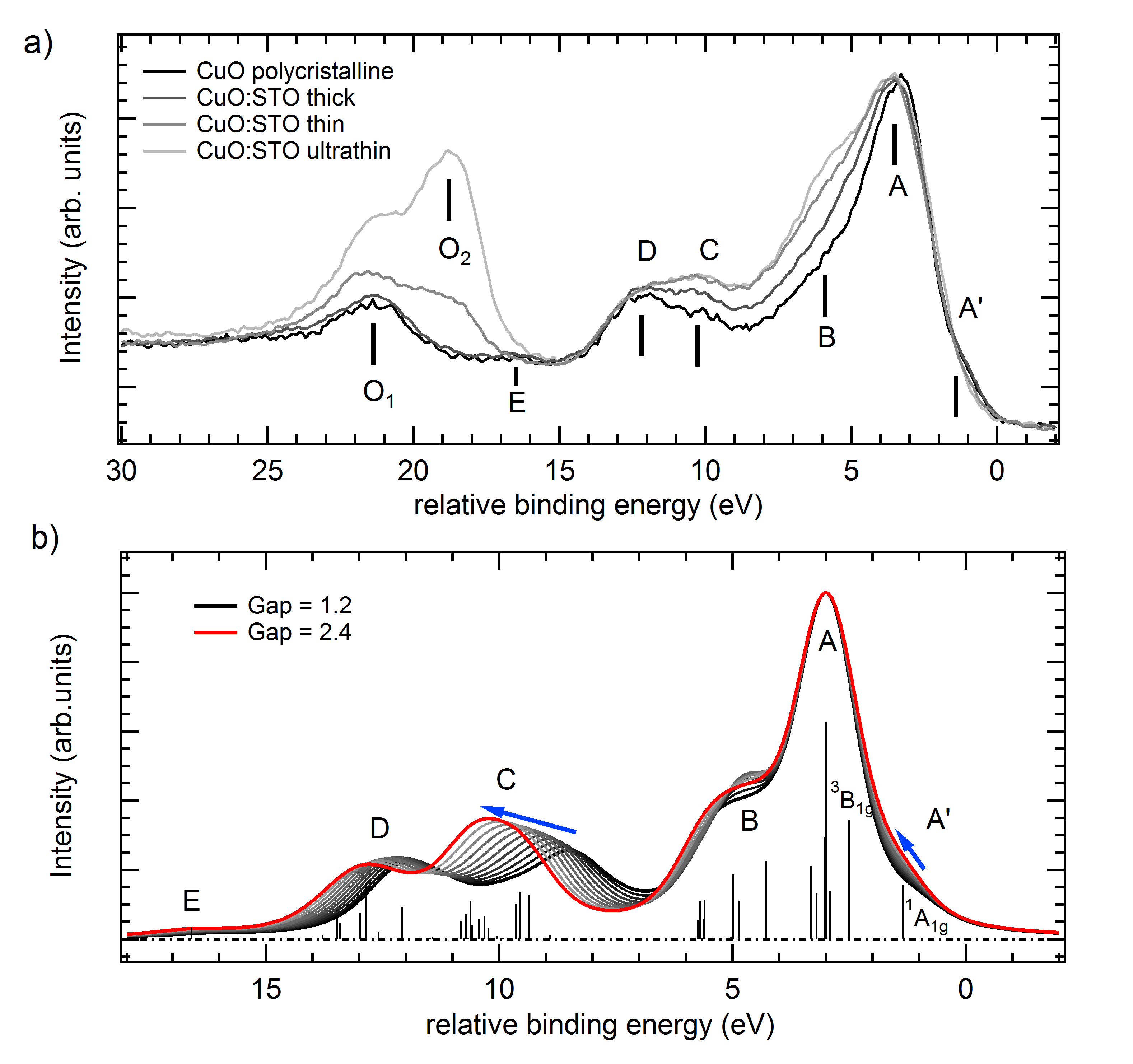}
\caption{a) experimental valence band spectra (lighter color is associated to thinner films); spectra have been aligned to Cu 3s contribution of polycrystalline spectra for a better comparison. b) calculated VB spectra from the cluster-model (described in Ref.\cite{Eskes1990}), for various band-gap. Bars refer to the case of E$_g$ = 2.4 eV. \label{fig_XPSVB_A}}
\end{center}
\end{figure}

Further details about the m-CuO and t-CuO electronic structure can be found in the valence band (VB) photoemission spectra, shown in Fig. \ref{fig_XPSVB_A}-a. In this figure, the VB have been aligned with the Cu 3s core line of polycristalline CuO in order to allow for a better comparison; please note that absolute valence band maxima (VBM) values have been obtained on VB spectra aligned with the C 1s reference peak. The O$_1$ and O$_2$ features correspond to the O 2s shallow core levels of CuO and STO, respectively; the substrate contribution can not be detected in the Thick CuO/STO due to the high overlayer thickness. The measured XPS valence band spectra, which span a 15 eV range, show a main line (A) with a shoulder on the high BE side (B), and a satellite with two features (C, D) in the 8-14 eV BE range.

The polycristalline m-CuO spectra are consistent with the literature, while a noticeable increase of C and B features (Fig. \ref{fig_XPSVB_A}) can be detected in the heterostructures, alongside with a small relative shift of the uppermost energy level (A'). It should be noted that a small change of spectral weight of C and B components is also observed on Thick CuO/STO sample; the overall VB spectra increase in thinner films can not be completely justified by the presence of STO VB contribution alone, which in any case would raise the photoelectron intensity only at the B position (see also Fig. \ref{fig_BandsCuO}-a).

As for the case of the 2p levels, the CuO valence band spectrum can be predicted with a configurational-interaction cluster-model calculation\cite{JELSPECParmigiani,Eskes1990}, by considering the crystal field effect in a distorted octahedral environment (D$_{4h}$ symmetry). Figure \ref{fig_XPSVB_A}-b shows the calculation results for several parameters, chosen to mimic a band-gap increase from 1.2 eV to 2.4 eV, as outlined in Table II of Ref.\cite{Eskes1990}. The blue arrows point out the major changes of spectral weight with the increasing band-gap, which follows the same trend of the experimental spectra. In fact, as the energy gap increases from 1.2 to 2.4 eV, the calculations show an increase of the B and C structures, along with a slight reduction of the A' shoulder. This feature was related to the Zhang-Rice singlet (see, e.g., Refs. \cite{Clemens2016,Hamad2018} and Refs. therein) and was recently mapped by ARPES\cite{Grioni2014}. The present results show that the A' feature, labeled as $^1$A$_{1g}$ singlet state according to the D$_{4h}$ cluster calculation, moves towards the second lowest lying state ($^3$B$_{1g}$ symmetry) as the band-gap increases.
The changes of the spectral weight in the VB region (especially C and A' features) with respect to m-CuO were also evidenced\cite{JELSPECParmigiani} in CuGeO$_3$ and Bi$_2$CuO$_4$, and mainly related to changes in the charge transfer energy.

\subsection{Band alignment and spectroscopic ellipsometry}
In order to obtain the conduction band minima positions, the value of the band-gap must be added to the VBM calculated from XPS data; yet, the reported values for CuO gap can vary in the 1.35 - 1.7 eV range\cite{DFT-AllCuOxides}, both from experimental and theoretical data. We then resorted to spectroscopic ellipsometry to directly probe the t-CuO gap.

\begin{figure}[h!]
\begin{center}
\includegraphics[width=0.50\textwidth]{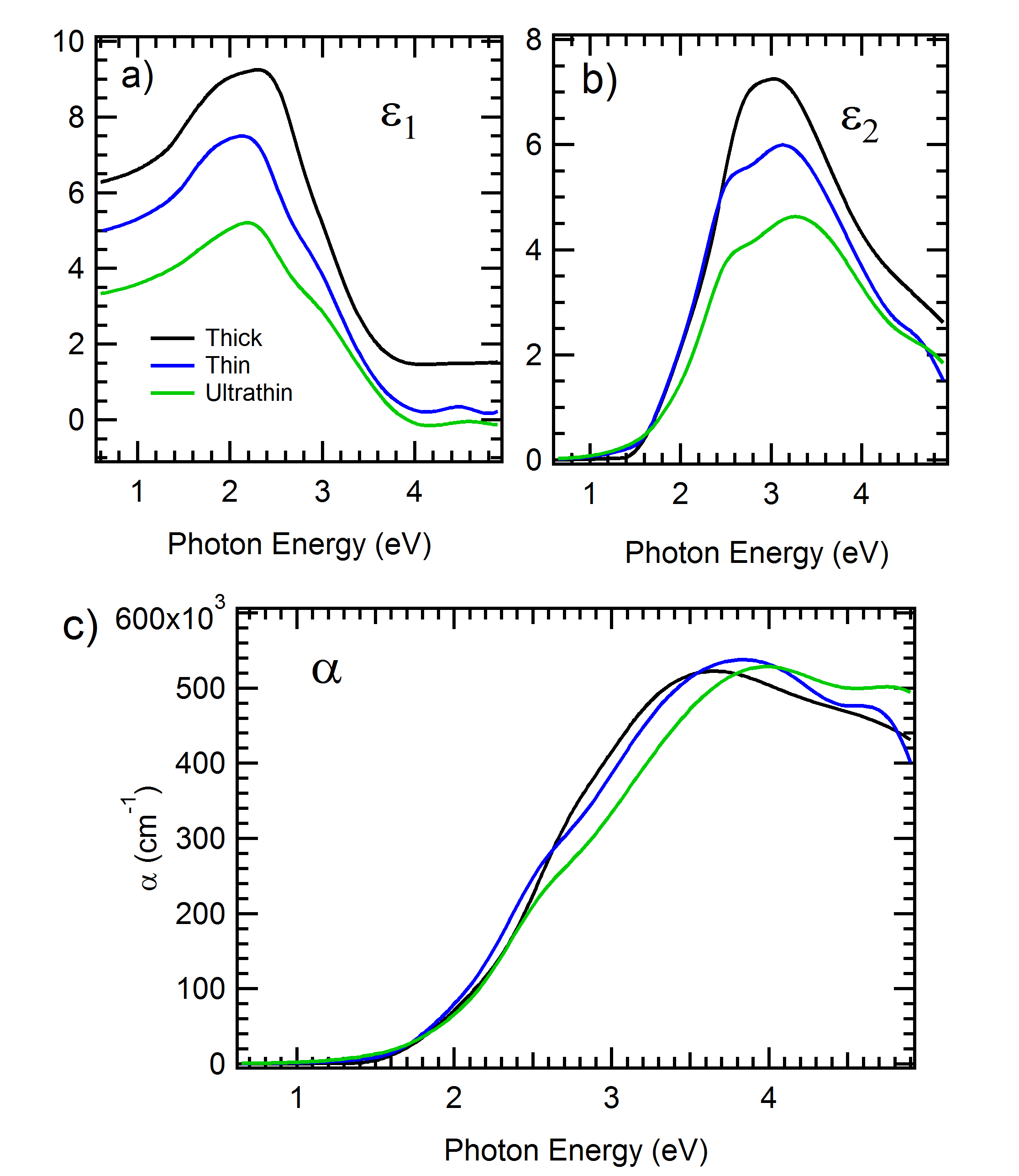}
\caption{The complex dielectric function $\epsilon$ components $\epsilon_1$ (a) and $\epsilon_2$ (b) of the CuO films as evaluated from SE. c) Absorption coefficient measured on the CuO/SrTiO$_3$ samples. \label{fig_Absorption}}
\end{center}
\end{figure}

In general, the ellipsometric analysis allows\cite{SE1}: i) to determine the dielectric function spectra of the film by best-fitting simulated spectra to the experimental ones; ii) to verify the correspondence between nominal and actual film thickness values; iii) to check the optical quality and uniformity of the films.

First, we characterized the dielectric function of SrTiO$_3$ substrate, which was in a good agreement with the database data. Then, the thin film samples were analyzed, adopting a three-phase model made of ambient, CuO film and SrTiO$_3$. The complex dielectric function $\epsilon = \epsilon_1 + i\epsilon_2$ (with $\epsilon = (n + i k)^2$, where $n$ is the refractive index and $k$ the extinction coefficient) of the CuO films was first argued from numerical inversion of SE data adopting the nominal film thicknesses, as provided by XPS analysis and growth rate estimation. The resulting values for the absorption coefficient  $\alpha$ ($\alpha = 4 \pi k/ \lambda$) indicate that the penetration depth of light in the film reaches tens of nanometers, so that the three-phase model has three unknown parameters: $\epsilon_1$, $\epsilon_2$ and the CuO thickness $d$.

We then performed a best-fit of the simulated spectra to the experimental ones. We modeled the dielectric function of CuO films with Tauc-Lorentz\cite{SE2} and Gaussian oscillators and obtained best-fitted $d$ values. The agreement between experimental and fitted spectra is very good and Kramers-Kronig consistency is guaranteed by the use of physical oscillators. Moreover, the actual thickness values are within 5\% to the nominal ones, as shown in Tab.\ref{tab_CuO}.

The dielectric function and absorption coefficient spectra (Fig. \ref{fig_Absorption}-a,b,c) slightly change with the film thickness, being somehow different for the ultrathin sample. This behavior is however compatible with the few material literature data and probably depends on the surface conditions. Though the nature of the band-gap is still subject of some debate\cite{MeyerCuOReviee}, in the present case we have determined for the Thick sample an 1.35 eV gap, in accordance to both experimental \cite{Koffyberg20} and theoretical studies \cite{Wu21}. With decreasing the film thickness the band-gap energy increases by 0.2 eV (i.e. to 1.57 eV), consistently with the previous XPS analysis; SE gap results are given in Tab. \ref{tab_CuO}.

The measured indirect gap values are in any case lower than our calculated DFT+U results for U $=$ 7.15 eV. Apart from the known limitations of DFT for gap calculations, such discrepancy could be also due to the strong electron-phonon coupling of CuO, which causes the reduction of the band-gap as the temperature increases. Such phenomenon has been observed by temperature variable absorption measurements by Marabelli et al.\cite{Marabelli19}, showing the trailing edge of absorbance data scaling down with the temperature; in that work, the gap values of 1.67 eV (consistent with ground-state DFT+U calculation) resulted in an absorbance trailing edge of nearly 1.35 eV (consistent to our experimental data) at room temperature, due to a strong electron-phonon coupling.

\begin{figure}[h!]
\begin{center}
\includegraphics[width=0.50\textwidth]{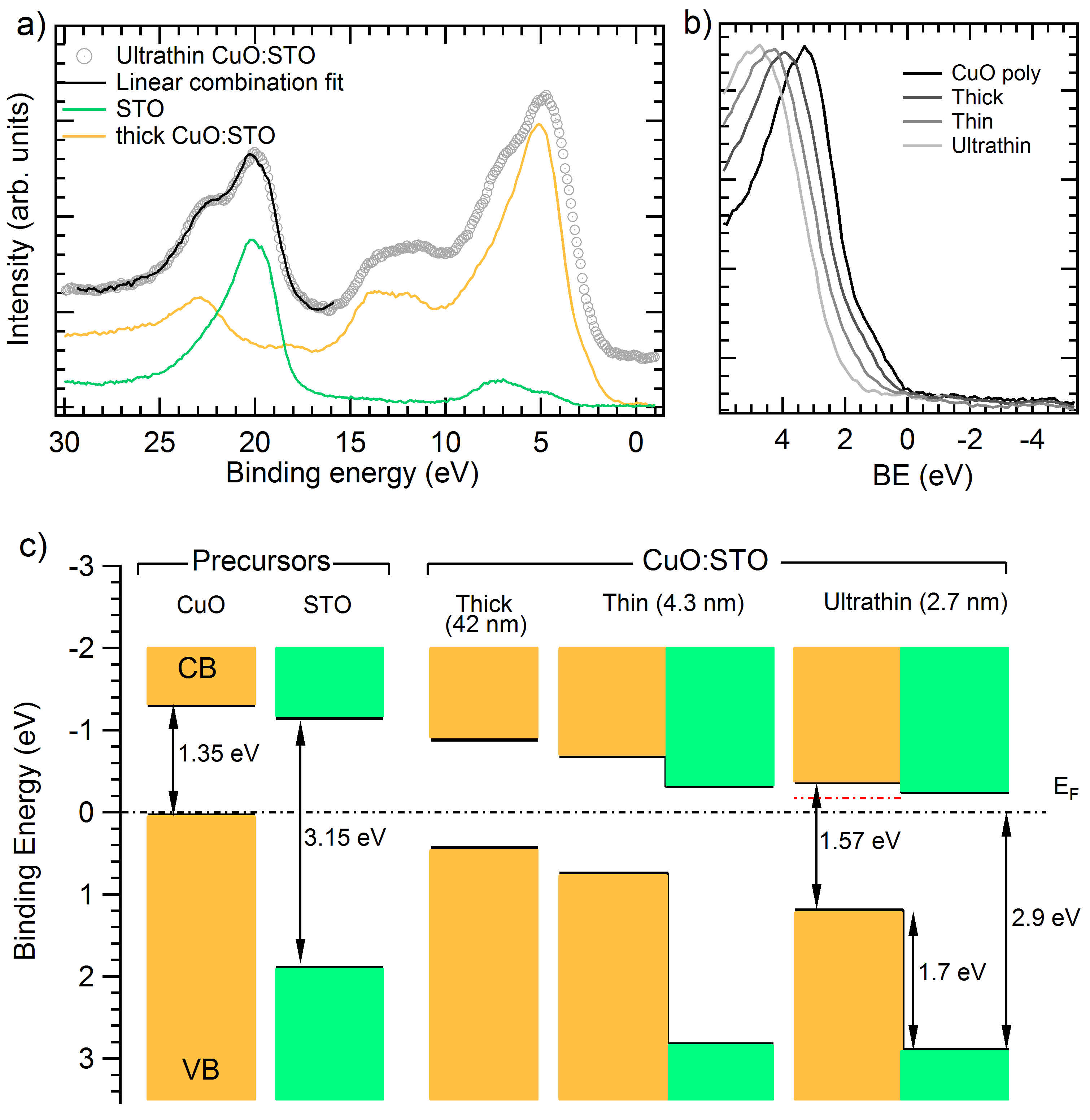}
\caption{XPS analysis of the band alignment. a): Linear combination of the CuO and SrTiO$_3$ signals (green and yellow, respectively), which produces the fit (black line) of the signal of the CuO/SrTiO$_3$ heterointerface; b): detail of VBM of all heterostructures and polycrystalline CuO; c): schematic band alignment results for heterostructures and for the bare CuO and STO precursor. For the Ultrathin film case, the position CBM evaluated from the bulk m-CuO bands is reported for comparison (red dashed lines). The band-gap values for CuO have been obtained by SE measurements.\label{fig_BandsCuO}}
\end{center}
\end{figure}

Finally, the XPS-derived band alignment at the CuO/SrTiO$_3$ interfaces are shown. The STO VBMs are obtained (for low CuO thickness) by fitting the valence band spectrum of the heterojunction with the linear combination of the spectra characteristic of pristine parent compounds\cite{Giampietri17}, which allows for an accurate alignment of their relative energy position (Figure \ref{fig_BandsCuO}-(a)). In particular, the shallow core levels at BE $\approx$20 eV have been used to accurately pinpoint the relative position of CuO and STO contribution in the junction (black line in Fig.\ref{fig_BandsCuO}-a). The CuO VBM can be directly evaluated from the data, shown in the actual binding energy scale in Fig.\ref{fig_BandsCuO}-b; a shift towards high BE is clearly detectable for thinner CuO films. As already pointed out, some limitations to this method come from the different VB shape for CuO, due to the different electronic structure of t-CuO and m-CuO. Combined to the intrinsic difficulties to accurately determine the VBM, the leading edge position displays an uncertainty of $\pm$ 0.15 eV, consistently with the evaluation shown by Chambers et al.\cite{Chambers_Leading_edge} for this method. The experimentally measured band schematics for heterostructures and separated precursors are shown in Fig.\ref{fig_BandsCuO}-c.

By using the measured values, the band alignment for the Thick and Thin films are thus characterized by a staggered, type II interface (Fig.\ref{fig_BandsCuO}-(b)); in such configuration, a charge confinement within the SrTiO$_3$ substrate could be possible, although t-CuO is a non-polar solid on (001) direction. By considering the 1.35 eV gap of m-CuO, in the Ultrathin case a Type I junction would be expected (red dashed line in Fig.\ref{fig_BandsCuO}-c). However, with the measured 0.2 eV band-gap increase, the junction changes into the type II configuration as well. By considering the gap temperature dependence\cite{Marabelli19}, the conduction band displacement is expected to significatively increase at lower temperatures.

With respect to the polycrystalline sample, the CuO VBM is shifted away from the Fermi level, up to 1.2 eV for the Ultrathin case. This shift, which corresponds to n-type doping, is rather unusual for CuO, which is usually considered analogous to p-type semiconductor. A strong shift of the STO bands towards the Fermi level with respect to the clean STO (001) surface is also observed, similar to the one reported for other perovskite-based heterostructures \cite{Giampietri18}. This shift is usually related to a strong n-type doping in STO, which in turn can be related to the presence of charge in STO whenever the band bending produces structures which cross the Fermi level. However, by considering the Ultrathin case, the nearly zero conduction bands offset at the CuO/SrTiO$_3$ junction is not compatible with an electron charge transfer in a preferential direction, at odds with other SrTiO$_3$-based heterostructures where a large CBO usually leads to a strong charge confinement within the substrate \cite{Giampietri17}.

The experimental junction diagram is then compatible with any charge-separation application (such as photovoltaic and photocatalysis), down to few unit cell thickness; in particular, these results are in very good agreement with the proposed diagram for the enhanced photoelectrochemical water splitting\cite{Choudhary6} already observed in CuO/STO heterostructures.

\section{Conclusions}
In conclusion, several epitaxial CuO thin films have been grown on SrTiO$_3$ by RF sputtering and characterized by AFM, XPD and SE analysis. Photoelectron spectroscopies (XPS and XPD) reveal the presence of CuO films with the proper stoichiometry, a nominal oxidation state of the Ti atoms in the substrate and the clear presence of the t-CuO phase for ultrathin films. A direct measure of the band-gap of CuO allowed us to reconstruct the band alignment at the interface. XPS and SE measurements are consistent with a gap increase of 0.2 eV in t-CuO (2.7 nm), leading to a type II junction instead of a type I. Upon excitation, the conduction band offset in the CuO/SrTiO$_3$ junction suggests a confinement of mobile charges (electrons) within the CuO film. Additionally, a strong shift of the Fermi level towards the CBM as a function of film thickness has been measured. For all the studied films an indirect band-gap is observed; the gap type and broadening results are also consistent with DFT+U calculations.

\bibliography{biblio}
\end{document}